\documentclass[aps,prl,reprint,superscriptaddress,preprintnumbers]{revtex4-2}
\usepackage[T1]{fontenc}
\usepackage{times}
\usepackage{graphicx,color}
\usepackage{amsfonts,amsmath,amssymb,amsbsy}
\usepackage[colorlinks=true, linkcolor=blue, citecolor=blue, urlcolor=blue]{hyperref}

\def\H{{\mathcal{H}}}
\let\mathbf=\boldsymbol

\def\blue#1{\textcolor{blue}{#1}}
\def\emph#1{\textcolor{magenta}{#1}}
\begin{document}

\title{Dynamic transformation between a skyrmion string and a bimeron string \\ in a layered frustrated system}

\author{Xichao Zhang}
\thanks{These authors contributed equally to this work.}
\affiliation{Department of Electrical and Computer Engineering, Shinshu University, 4-17-1 Wakasato, Nagano 380-8553, Japan}

\author{Jing Xia}
\thanks{These authors contributed equally to this work.}
\affiliation{School of Science and Engineering, The Chinese University of Hong Kong, Shenzhen, Guangdong 518172, China}

\author{Oleg A. Tretiakov}
\affiliation{School of Physics, The University of New South Wales, Sydney 2052, Australia}

\author{Hung T. Diep}
\affiliation{Laboratoire de Physique Th{\'e}orique et Mod{\'e}lisation, CY Cergy Paris Universit{\'e}, 95302 Cergy-Pontoise Cedex, France}

\author{Guoping Zhao}
\affiliation{College of Physics and Electronic Engineering, Sichuan Normal University, Chengdu 610068, China}

\author{Jinbo Yang}
\affiliation{State Key Laboratory for Mesoscopic Physics, School of Physics, Peking University, Beijing, 100871, China}

\author{\\ Yan Zhou}
\email[]{zhouyan@cuhk.edu.cn}
\affiliation{School of Science and Engineering, The Chinese University of Hong Kong, Shenzhen, Guangdong 518172, China}

\author{Motohiko Ezawa}
\email[]{ezawa@ap.t.u-tokyo.ac.jp}
\affiliation{Department of Applied Physics, The University of Tokyo, 7-3-1 Hongo, Tokyo 113-8656, Japan}

\author{Xiaoxi Liu}
\email[]{liu@cs.shinshu-u.ac.jp}
\affiliation{Department of Electrical and Computer Engineering, Shinshu University, 4-17-1 Wakasato, Nagano 380-8553, Japan}

\begin{abstract}
Frustrated topological spin textures have unique properties that may enable novel spintronic applications, such as helicity-based information storage and computing. Here, we report the statics and current-induced dynamics of two-dimensional (2D) pancake skyrmions in a stack of weakly coupled frustrated magnetic monolayers, which form a three-dimensional (3D) skyrmion string. The Bloch-type skyrmion string is energetically more stable than its N{\'e}el-type counterpart. It can be driven into translational motion by the dampinglike spin-orbit torque and shows the damping-dependent skyrmion Hall effect. Most notably, the skyrmion string can be transformed to a dynamically stable bimeron string by the dampinglike spin-orbit torque. The current-induced bimeron string rotates stably with respect to its center, which can spontaneously transform back to a skyrmion string when the current is switched off. Our results reveal unusual physical properties of 3D frustrated spin textures, and may open up different possibilities for spintronic applications based on skyrmion and bimeron strings.
\end{abstract}

\date{December 16, 2021}

\preprint{\href{https://doi.org/10.1103/PhysRevB.104.L220406}{\blue{PHYSICAL REVIEW B \textbf{104}, L220406 (2021)}}}


\maketitle


\textit{Introduction.}
Topological spin textures are particlelike objects that can be used as robust information carriers for data processing~\cite{Roszler_NATURE2006,Muhlbauer_SCIENCE2009,Yu_Nature2010,Nagaosa_NNANO2013,Mochizuki_Review,Wiesendanger_Review2016,Finocchio_JPD2016,Kang_PIEEE2016,Kanazawa_AM2017,Wanjun_PHYSREP2017,Fert_NATREVMAT2017,Everschor_JAP2018,Zhou_NSR2018,Zhang_JPCM2020,Gobel_PP2021}.
Frustrated spin systems can host different species of topological spin textures~\cite{Okubo_PRL2012,Leonov_NCOMMS2015,Lin_PRB2016A,Hayami_PRB2016A,Rozsa_PRL2016,Leonov_NCOMMS2017,Kharkov_PRL2017,Xichao_NCOMMS2017,Yuan_PRB2017,Hou_AM2017,Hu_SR2017,Malottki_SR2017,Liang_NJP2018,Ritzmann_NE2018,Kurumaji_SCIENCE2019,Desplat_PRB2019,Xia_PRApplied2019,Zarzuela_PRB2019,Lohani_PRX2019,Gobel_PRB2019,Diep_PRB2019,Diep_Symmetry2020,Diep_Entropy2019,Zhang_PRB2020,Xia_APL2020,Zhang_APL2021,Batista_2016,Psaroudaki_PRL2021,Sinova_2021A,Sinova_2021B}, which show very different physical properties and behaviors compared to their common ferromagnetic (FM) counterparts.
For example, skyrmions carrying different topological charges, either positive or negative, can be stabilized in a perpendicularly magnetized monolayer with exchange frustration~\cite{Okubo_PRL2012,Leonov_NCOMMS2015,Lin_PRB2016A,Hayami_PRB2016A,Rozsa_PRL2016,Leonov_NCOMMS2017,Kharkov_PRL2017,Xichao_NCOMMS2017,Yuan_PRB2017,Hou_AM2017,Hu_SR2017,Malottki_SR2017,Liang_NJP2018,Ritzmann_NE2018,Kurumaji_SCIENCE2019,Desplat_PRB2019,Xia_PRApplied2019,Zarzuela_PRB2019,Lohani_PRX2019,Gobel_PRB2019,Diep_PRB2019,Diep_Symmetry2020,Diep_Entropy2019,Xia_APL2020,Batista_2016}.
In contrast, skyrmions in common chiral magnets are stabilized by Dzyaloshinskii-Moriya (DM) exchange interactions~\cite{Roszler_NATURE2006,Muhlbauer_SCIENCE2009,Yu_Nature2010,Nagaosa_NNANO2013,Mochizuki_Review,Wiesendanger_Review2016,Finocchio_JPD2016,Kang_PIEEE2016,Kanazawa_AM2017,Wanjun_PHYSREP2017,Fert_NATREVMAT2017,Everschor_JAP2018,Zhou_NSR2018,Zhang_JPCM2020,Gobel_PP2021,Wanjun_NPHYS2017,Litzius_NPHYS2017}, and skyrmions with large or negative topological charges are usually unstable in chiral magnets with symmetric DM interactions~\cite{Zhang_PRB2016}.
Other exemplary topological spin textures in frustrated spin systems include the so-called skyrmioniums~\cite{Xia_APL2020}, bimerons~\cite{Zhang_PRB2020,Kharkov_PRL2017,Gobel_PRB2019,Sinova_2020}, and bimeroniums~\cite{Zhang_APL2021}, all of which are functional building blocks for spintronic applications~\cite{Finocchio_JPD2016,Kang_PIEEE2016,Fert_NATREVMAT2017,Everschor_JAP2018,Zhou_NSR2018,Zhang_JPCM2020,Gobel_PP2021,Sinova_2017}.

Recent studies on frustrated skyrmions have mainly focused on the static and dynamic properties of frustrated skyrmions in two-dimensional (2D) space~\cite{Okubo_PRL2012,Leonov_NCOMMS2015,Lin_PRB2016A,Hayami_PRB2016A,Rozsa_PRL2016,Leonov_NCOMMS2017,Kharkov_PRL2017,Xichao_NCOMMS2017,Yuan_PRB2017,Hou_AM2017,Hu_SR2017,Malottki_SR2017,Liang_NJP2018,Ritzmann_NE2018,Kurumaji_SCIENCE2019,Desplat_PRB2019,Xia_PRApplied2019,Zarzuela_PRB2019,Lohani_PRX2019,Gobel_PRB2019,Diep_PRB2019,Diep_Symmetry2020,Diep_Entropy2019,Zhang_PRB2020,Xia_APL2020,Zhang_APL2021,Batista_2016,Psaroudaki_PRL2021}.
In particular, theoretical works have shown that the helicity dynamics of a 2D frustrated skyrmion is coupled to its center-of-mass dynamics~\cite{Leonov_NCOMMS2015,Lin_PRB2016A,Leonov_NCOMMS2017,Xichao_NCOMMS2017,Ritzmann_NE2018,Xia_PRApplied2019,Zhang_PRB2020}.
This property is in stark contrast to that of 2D skyrmions stabilized in chiral magnets, where the helicity of a moving skyrmion is strictly locked by the DM exchange interaction~\cite{Roszler_NATURE2006,Muhlbauer_SCIENCE2009,Yu_Nature2010,Nagaosa_NNANO2013,Mochizuki_Review,Wiesendanger_Review2016,Finocchio_JPD2016,Kang_PIEEE2016,Kanazawa_AM2017,Wanjun_PHYSREP2017,Fert_NATREVMAT2017,Everschor_JAP2018,Zhou_NSR2018,Zhang_JPCM2020,Gobel_PP2021}.
This feature also implies that 2D frustrated skyrmions have more degrees of freedom that, in principle, can be manipulated by external stimuli and used for building future spintronic devices~\cite{Leonov_NCOMMS2015,Leonov_NCOMMS2017,Xichao_NCOMMS2017,Xia_PRApplied2019,Zhang_PRB2020,Xia_APL2020,Zhang_APL2021,Psaroudaki_PRL2021}.
For example, several studies have suggested that the information can be encoded by the location of skyrmions with unity topological charges in chiral magnets~\cite{Finocchio_JPD2016,Kang_PIEEE2016,Fert_NATREVMAT2017,Everschor_JAP2018,Zhou_NSR2018,Zhang_JPCM2020,Gobel_PP2021}.
In frustrated spin systems, the information can be carried by the topological charge of skyrmions or be encoded by the helicity of skyrmions~\cite{Leonov_NCOMMS2015,Leonov_NCOMMS2017,Xichao_NCOMMS2017,Xia_PRApplied2019,Zhang_PRB2020,Xia_APL2020,Zhang_APL2021,Psaroudaki_PRL2021}.

However, the physical properties and potential applications of frustrated skyrmions in three-dimensional (3D) structures still remain elusive and thus represent an area of significant opportunity for research.
As an analogy to the 3D vortex line formed by 2D pancake vortices~\cite{Clem_1991,Clem_2004} in a stack of coupled superconducting layers~\cite{Reichhardt_PRB2001,Liu_APL2020}, a 3D frustrated skyrmion string can be constructed by a stack of pancake skyrmions in a frustrated multilayer~\cite{Lin_PRL2018,Reichhardt_2021}, where each frustrated pancake skyrmion is a 2D object.
Such a 3D skyrmion string is an important component for future spintronic applications based on 3D nanostructures~\cite{Lin_PRL2018,Reichhardt_2021} and layered systems~\cite{Reichhardt_PRB2011,Xia_PRB2021,Reichhardt_2021}.
In this Letter, we report the statics and dynamics of such a stack of frustrated pancake skyrmions, where the pancake skyrmions in two adjacent monolayer are coupled via a FM interlayer exchange coupling.

\begin{figure*}[t]
\centerline{\includegraphics[width=0.99\textwidth]{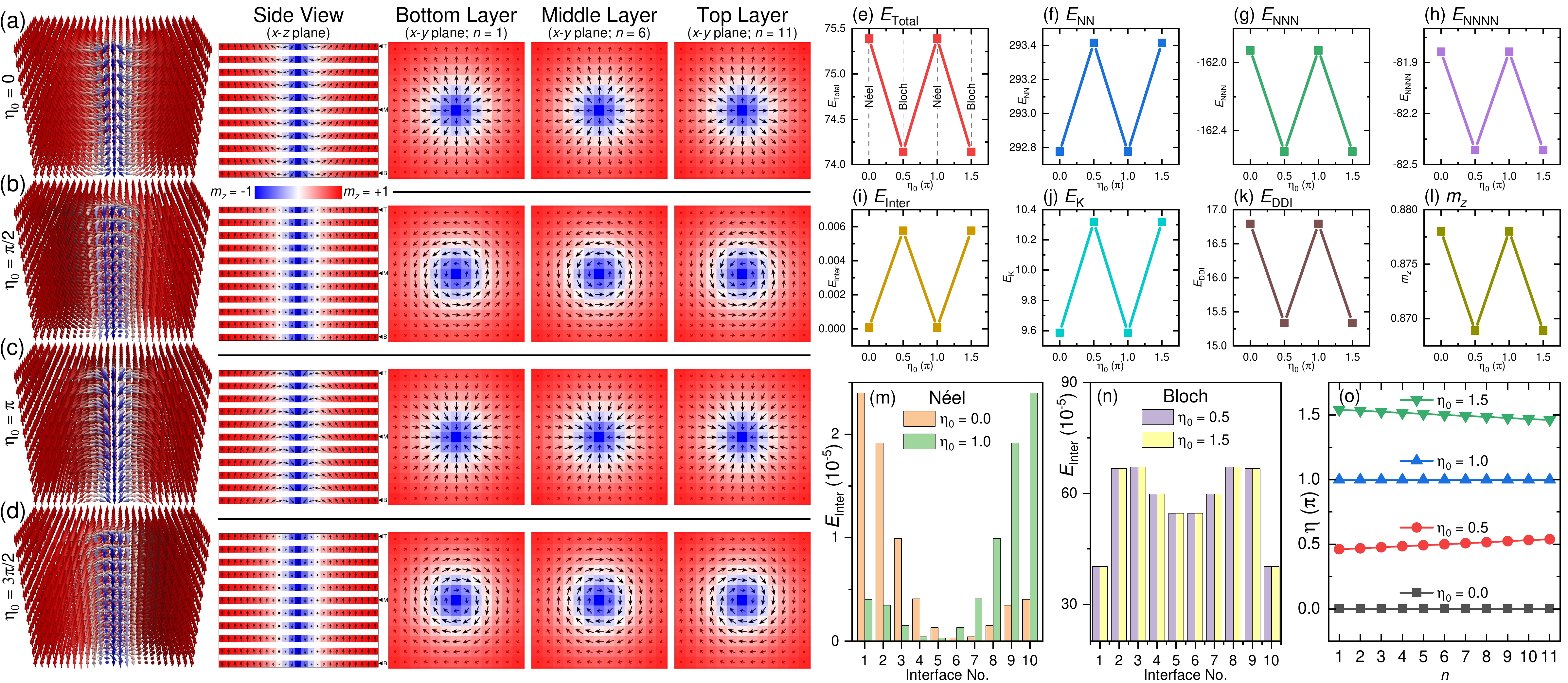}}
\caption{%
3D and 2D illustrations of static skyrmion strings that are relaxed with the initial helicity of (a) $\eta_0=0$, (b) $\eta_0=\pi/2$, (c) $\eta_0=\pi$, and (d) $\eta_0=3\pi/2$.
The 3D and 2D side views show the vertical cross sections through the core of the skyrmion string. The 2D top views show the horizontal cross sections through the bottommost ($n=1$), middle ($n=6$), and topmost ($n=11$) FM layers, and focus on the skyrmion core area. The arrow represents the spin direction. The color scale represents the out-of-plane spin component $m_z$, which has been used throughout this Letter.
(e) Total energy $E_{\text{Total}}$, (f) NN exchange energy $E_{\text{NN}}$, (g) NNN exchange energy $E_{\text{NNN}}$, (h) NNNN exchange energy $E_{\text{NNNN}}$, (i) total interlayer exchange energy $E_{\text{inter}}$, (j) PMA energy $E_{\text{K}}$, (k) DDI energy $E_{\text{DDI}}$, and (l) $m_z$ as functions of $\eta_0$ are given.
All energies are given in units of $J_{1}=1$.
Interlayer exchange energies as functions of interface number are given for the skyrmion strings relaxed with (m) $\eta_0=0,1$ and (n) $\eta_0=\pi/2,3\pi/3$.
(o) Layer-dependent helicity $\eta$ of the relaxed skyrmion strings.
}
\label{FIG1}
\end{figure*}

\begin{figure*}[t]
\centerline{\includegraphics[width=0.99\textwidth]{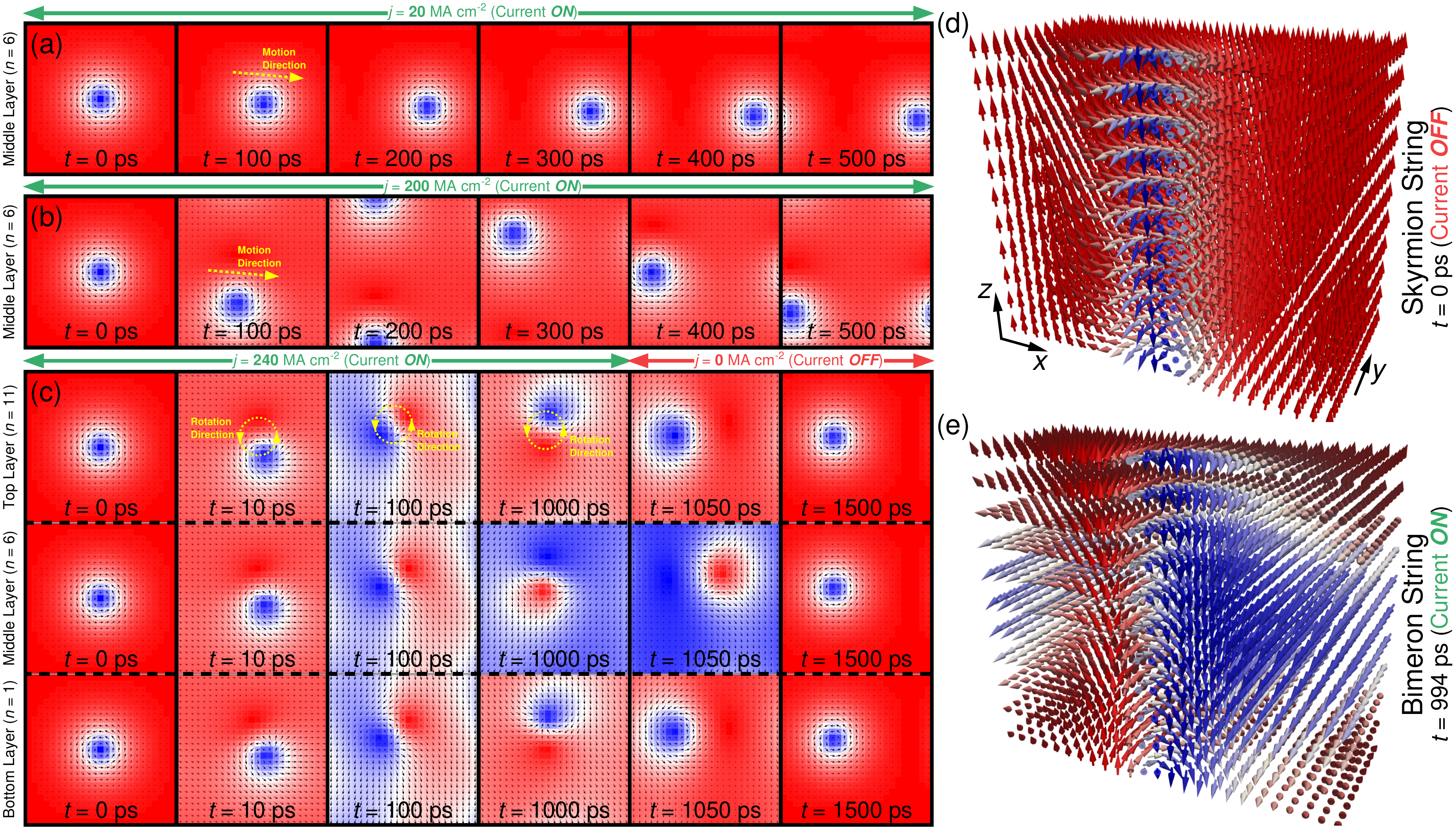}}
\caption{%
Top views of a Bloch-type skyrmion string driven by (a) a small current $j=20$ MA cm$^{-2}$ and (b) a large current $j=200$ MA cm$^{-2}$.
The spin configurations are similar in all FM layers, so only the spin configuration of the middle layer ($n=6$) is given.
(c) Top views of the current-controlled mutual transformation between a skyrmion string and a bimeron string.
$j=240$ MA cm$^{-2}$ is applied for $t=0-1000$ ps, followed by a $500$-ps-long relaxation. The spin configurations of the bottommost ($n=1$), middle ($n=6$), and topmost ($n=11$) layers are given.
(d) 3D view of the core of the skyrmion string at $t=0$ ps.
(e) 3D view of the core of the bimeron string at $t=995$ ps.
}
\label{FIG2}
\end{figure*}

\begin{figure*}[t]
\centerline{\includegraphics[width=0.99\textwidth]{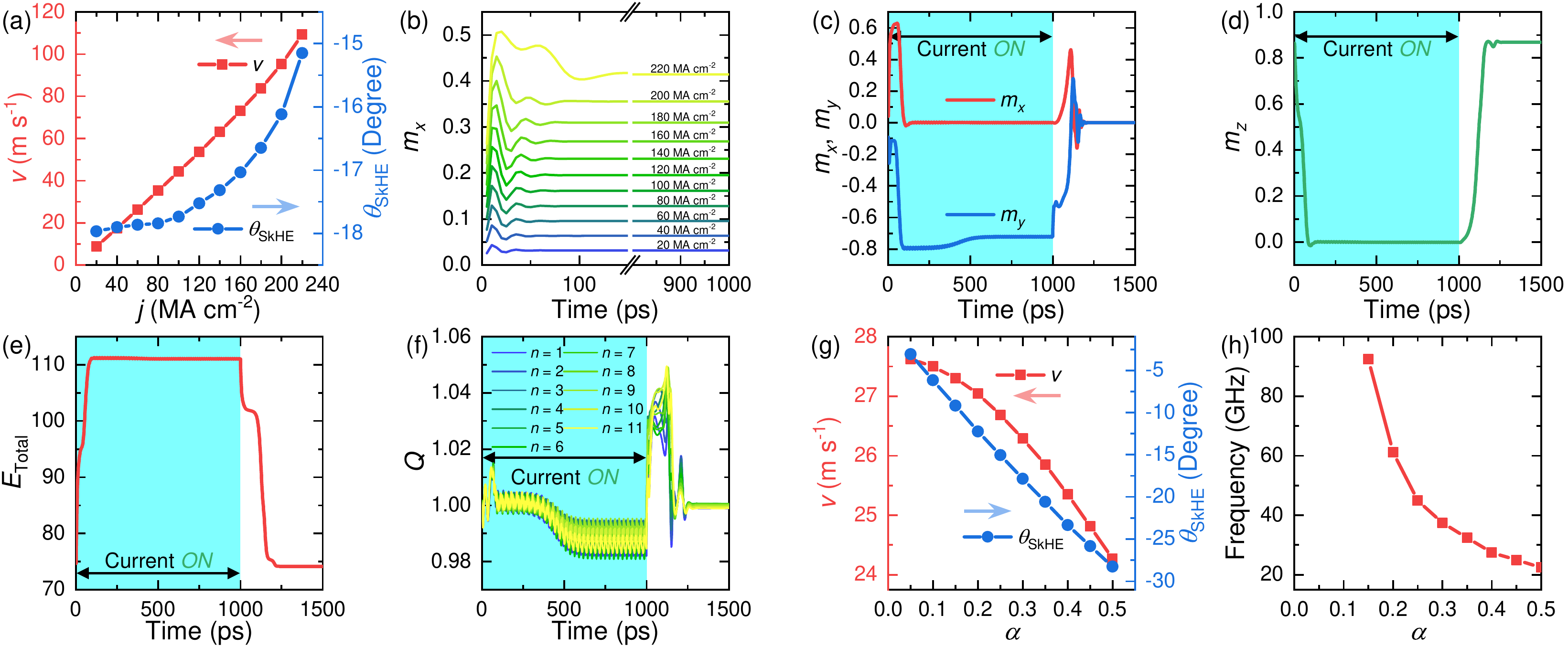}}
\caption{%
(a) Velocity $v$ and skyrmion Hall angle $\theta_{\text{SkHE}}$ as functions of $j$ for a Bloch-type skyrmion string. 
(b) In-plane spin component $m_{x}$ as a function of time at different $j$, corresponding to (a).
(c) $m_{x,y}$, (d) $m_{z}$, (e) $E_{\text{Total}}$, and (f) $n$-dependent absolute topological charge $\left|Q\right|$ as functions of time for the current-controlled mutual transformation between a skyrmion string and a bimeron string, where $j=240$ MA cm$^{-2}$ is applied for $t=0-1000$ ps.
(g) $\alpha$-dependent $v$ and $\theta_{\text{SkHE}}$ of a Bloch-type skyrmion string driven by $j=60$ MA cm$^{-2}$.
(h) $\alpha$-dependent rotation frequency of a bimeron string driven by $j=240$ MA cm$^{-2}$.
}
\label{FIG3}
\end{figure*}


\textit{Model.}
To be specific, we consider a 3D skyrmion string formed by $11$ aligned stacks of 2D pancake skyrmions in a frustrated spin system.
Each 2D FM layer has $25\times 25$ spins and is described by a $J_{1}$-$J_{2}$-$J_{3}$ classical Heisenberg model on a simple square lattice~\cite{Kaul_2004,Lin_PRB2016A,Xichao_NCOMMS2017,Xia_PRApplied2019,Zhang_PRB2020,Xia_APL2020,Diep_Entropy2019,Batista_2016}, of which the Hamiltonian $\H_{n}$ reads
\begin{align}
\H_n=&-J_1\sum_{\substack{<i,j>}}\boldsymbol{m}_i^n\cdot\boldsymbol{m}_j^n-J_2\sum_{\substack{\ll i,j\gg}}\boldsymbol{m}_i^n\cdot\boldsymbol{m}_j^n \\
&-J_3\sum_{\substack{\lll i,j\ggg}}\boldsymbol{m}_i^n\cdot\boldsymbol{m}_j^n-K\sum_{\substack{i}}{(m^{n,z}_i)^{2}}+H_{\text{DDI}}, \notag
\label{eq:Hamiltonian_Intralayer}
\end{align}
where $n$ is the FM layer index ($n=1,2,\cdots,11$), and $\boldsymbol{m}_{i}^n$ represents the normalized spin at the site $i$ of layer $n$, $|\boldsymbol{m}_{i}^n|=1$.
$J_1$, $J_2$, and $J_3$ denote the FM nearest-neighbor (NN), antiferromagnetic (AFM) next-NN (NNN), and AFM next-NNN (NNNN) intralayer exchange interaction constants, respectively.
$\left\langle i,j\right\rangle$,
$\left\langle\left\langle i,j\right\rangle\right\rangle$, and
$\left\langle\left\langle\left\langle i,j\right\rangle\right\rangle\right\rangle$
run over all the NN, NNN, and NNNN sites in each FM layer, respectively.
$K$ is the perpendicular magnetic anisotropy (PMA) constant.
$H_{\text{DDI}}$ stands for the dipole-dipole interaction (DDI).

In our model, two NN FM layers are separated by a nonmagnetic heavy-metal spacer layer, which is required for realizing the interlayer coupling and spin current~\cite{Sinova_2015}.
We note that the spacers may consist of different heavy metals to ensure a net spin current.
The Hamiltonian $\H_{\text{inter}}$ for the interlayer coupling reads
\begin{equation}
\H_{\text{inter}}=-\sum_{n=1}^{10}J_{\text{inter}}\sum_{i}\boldsymbol{m}_{i}^{n}\cdot\boldsymbol{m}_{i}^{n+1}.
\label{eq:Hamiltonian_Interlayer}
\end{equation}
Hence, the total Hamiltonian of the system is written as $\H=\sum_{n=1}^{11}\H_{n}+\H_{\text{inter}}$.
We assume that the adjacent FM layers are coupled through a weak FM interlayer coupling $J_{\text{inter}}=0.01$ (in units of $J_1=1$).
We also assume that the spin dynamics is induced by the dampinglike spin-orbit torque $\tau_{\text{d}}$, which is described by the Landau-Lifshitz-Gilbert equation augmented with $\tau_{\text{d}}$~\cite{OOMMF},
\begin{equation}
\frac{d\boldsymbol{m}}{dt}=-\gamma_{0}\boldsymbol{m}\times\boldsymbol{h}_{\rm{eff}}+\alpha\left(\boldsymbol{m}\times\frac{d\boldsymbol{m}}{dt}\right)+\tau_{\text{d}},
\label{eq:LLG}
\end{equation}
where $\boldsymbol{h}_{\rm{eff}}=-\frac{1}{\mu_{0}M_{\text{S}}}\cdot\frac{\delta\mathcal{H}}{\delta\boldsymbol{m}}$ is the effective field,
$\mu_{0}$ is the vacuum permeability constant,
$M_{\text{S}}$ is the saturation magnetization,
$t$ is the time,
$\alpha$ is the Gilbert damping parameter,
and $\gamma_0$ is the absolute gyromagnetic ratio.
$\tau_{\text{d}}=\frac{u}{b}\left(\boldsymbol{m}\times\boldsymbol{p}\times\boldsymbol{m}\right)$
with $u=\left|\left(\gamma_{0}\hbar/\mu_{0}e\right)\right|\cdot\left(j\theta_{\text{SH}}/2 M_{\text{S}}\right)$ being the spin-torque coefficient.
$\hbar$ is the reduced Planck constant, $e$ is the electron charge, $b$ is the FM layer thickness, $j$ is the current density, and $\theta_{\text{SH}}$ is the spin Hall angle.
$\boldsymbol{p}=-\hat{y}$ denotes the spin polarization orientation.

The default parameters are~\cite{Lin_PRB2016A,Xichao_NCOMMS2017,Xia_PRApplied2019,Zhang_PRB2020,Xia_APL2020}:
$J_1=30$ meV,
$J_2=-0.8$ (in units of $J_1=1$),
$J_3=-0.6$ (in units of $J_1=1$),
$K=0.01$ (in units of $J_{1}/a^{3}=1$),
$\alpha=0.3$,
$\gamma_0=2.211\times 10^{5}$ m A$^{-1}$ s$^{-1}$,
$\theta_{\text{SH}}=0.2$,
and $M_{\text{S}}=580$ kA m$^{-1}$.
The lattice constant is $a=0.4$ nm. The mesh size is $a^3$.
We use the object oriented micromagnetic framework (\textsc{OOMMF})~\cite{OOMMF} upgraded with our extension modules to simulate the model.
We have simulated the metastability diagram using the OOMMF minimizer, which shows that the frustrated skyrmion strings are a metastable state for a wide range of $J_2$ and $J_3$ (see Ref.~\onlinecite{SI}). The minimum required value of $J_3$ for stabilizing a skyrmion string decreases with increasing $J_2$.


\textit{Static structures.}
We begin with simulating the static structure of a stack of coupled frustrated 2D pancake skyrmions in the absence of a driving current.
The interlayer coupling between adjacent pancake skyrmions leads to a 3D skyrmion string (Fig.~\ref{FIG1}).
The static structure of each pancake skyrmion is described by the topological charge
$Q={\frac{1}{4\pi}}\int\boldsymbol{m}(\boldsymbol{r})\cdot\left[\partial_{x}\boldsymbol{m}(\boldsymbol{r})\times\partial_{y}\boldsymbol{m}(\boldsymbol{r})\right]d^{2}\boldsymbol{r}$.
We parametrize each pancake skyrmion as
$\boldsymbol{m}(\boldsymbol{r})=\boldsymbol{m}(\theta,\phi)=(\sin\theta\cos\phi,\sin\theta\sin\phi,\cos\theta)$,
where we define
$\phi=Q_{\text{v}}\varphi+\eta$
with $\varphi$ being the azimuthal angle ($0\le\varphi<2\pi$).
Hence, $Q_{\text{v}}=\frac{1}{2\pi}\oint_{C}d \phi$ is the skyrmion vorticity and $\eta\in [0,2\pi)$ is the skyrmion helicity defined mod $2\pi$.
We assume that $Q_{\text{v}}=+1$ (i.e., $Q=-1$) and $\theta$ rotates by an angle of $\pi$ for spins from the skyrmion center to the skyrmion edge~\cite{Roszler_NATURE2006,Nagaosa_NNANO2013,Zhang_JPCM2020,Gobel_PP2021}.

The relaxed skyrmion strings consisting of N{\'e}el-type ($\eta=0,\pi$) or Bloch-type ($\eta=\pm\pi/2$) pancake skyrmions are given in Fig.~\ref{FIG1}.
Before the relaxation, a skyrmion with an initial helicity $\eta_0=0,\pi/2,\pi,3\pi/2$ is placed at the center of each FM layer. $\eta_0$ is identical in all FM layers.
Then, as shown in Figs.~\ref{FIG1}(a)-\ref{FIG1}(d), the N{\'e}el-type skyrmion strings with $\eta_0=0,\pi$ are relaxed to states with $\eta=\eta_0$ in each FM layer, while the Bloch-type skyrmion strings with $\eta_0=\pi/2,3\pi/2$ are relaxed to states with slightly nonuniform $\eta\sim\eta_0$ in each FM layer [Fig.~\ref{FIG1}(o)].
The total energies of the relaxed N{\'e}el-type skyrmion strings are larger than that of the Bloch-type skyrmion strings [Fig.~\ref{FIG1}(e)], indicating the N{\'e}el-type skyrmion strings are unstable states, largely due to the fact that the Bloch-type structures with $\eta=\pi/2,3\pi/2$ are favored by the DDI [Fig.~\ref{FIG1}(k)].
In general, the relaxed Bloch-type skyrmion strings have a slightly smaller out-of-plane magnetization [Fig.~\ref{FIG1}(l)], smaller NNN exchange [Fig.~\ref{FIG1}(g)], and NNNN exchange energies [Fig.~\ref{FIG1}(h)]. However, their NN exchange [Fig.~\ref{FIG1}(f)], interlayer exchange [Fig.~\ref{FIG1}(i)], and anisotropy energies [Fig.~\ref{FIG1}(j)] are slightly larger than that of relaxed N{\'e}el-type skyrmion strings.

The interlayer coupling energy is found to have a layer dependence for both relaxed N{\'e}el-type and Bloch-type skyrmion strings.
For the N{\'e}el-type skyrmion string with $\eta=0$ [Fig.~\ref{FIG1}(m)], the layer-dependent interlayer coupling energy reaches its maximum magnitude at the bottommost interface (i.e., the interface between $n=1$ and $n=2$).
For the N{\'e}el-type skyrmion string with $\eta=\pi$, the layer-dependent interlayer coupling energy reaches its maximum magnitude at the topmost interface (i.e., the interface between $n=10$ and $n=11$).
In contrast, for the Bloch-type skyrmion strings with $\eta\sim\pi/2,3\pi/2$, the layer-dependent interlayer coupling energy shows an identical M-profile dependence on the interfaces [Fig.~\ref{FIG1}(n)].
The interlayer coupling energy of the Bloch-type skyrmion string is larger than that of the N{\'e}el-type one, which is due to the slightly different in-plane spin configuration of each FM layer, as can be seen from the $n$-dependent $\eta$ in Fig.~\ref{FIG1}(o).
The $n$-dependent $\eta$ in the relaxed Bloch-type skyrmion string is caused by the DDI, which most commonly affects the in-plane spin configurations of the topmost ($n=11$) and bottommost ($n=1$) layers [Figs.~\ref{FIG1}(b) and~\ref{FIG1}(d)].
%

\textit{Current-induced dynamics.}
We further study the current-induced dynamics of a Bloch-type skyrmion string with $\eta\sim\pi/2$, which is initially relaxed at the sample center before the application of a driving current.
The sample include $11$ coupled FM layers with periodic boundary conditions in the $x$ and $y$ dimensions.
We first apply a current with a current density $j$ ranging from $20$ to $300$ MA cm$^{-2}$ to drive the pancake skyrmion in each FM layer. The effect of $\tau_{\text{d}}$ leads to the linear motion of the Bloch-type skyrmion string when $j=20-220$ MA cm$^{-2}$ (see Video 1 in Ref.~\onlinecite{SI}).

At a relatively smaller $j$, the skyrmion string moves stably and shows the skyrmion Hall effect [Fig.~\ref{FIG2}(a)], which is a natural consequence of the skyrmion Hall effect of the pancake skyrmion in each FM layer.
The variation of the skyrmion string in the $z$ dimension is very small during its steady motion, namely, there is almost no layer-dependent deformation in the skyrmion string. Hence, we calculate the skyrmion velocity and skyrmion Hall angle based on the skyrmion in the middle FM layer ($n=6$).
The skyrmion string velocity and its skyrmion Hall angle increase with $j$ when $j=20-220$ MA cm$^{-2}$ [Fig.~\ref{FIG3}(a)].
The change of the skyrmion Hall angle is due to the current-induced deformation of the skyrmion string, which can be seen from the selected top-view snapshots at $j=200$ MA cm$^{-2}$ [Fig.~\ref{FIG2}(b)] and $j$-dependent $m_x$-$t$ relation [Fig.~\ref{FIG3}(b)].

However, when $j\geq 240$ MA cm$^{-2}$, the skyrmion string smoothly transforms to a bimeron string when the current is applied [Figs.~\ref{FIG2}(c)-\ref{FIG2}(e)].
The current-induced formation of the bimeron string is due to the fact that the effect of $\tau_{\text{d}}$ with $\boldsymbol{p}=-\hat{y}$ tends to drag the spins in each FM layer from the $\pm z$ direction to the in-plane $-y$ direction [Figs.~\ref{FIG3}(c) and~\ref{FIG3}(d)].
Note that the bimeron in the in-plane magnetized system is a topological counterpart of the skyrmion in the perpendicularly magnetized system~\cite{Zhang_PRB2020}.
Once the bimeron string is formed under the driving current, it shows counter-clockwise rotation with a constant frequency determined by $j$ (see Videos 2-4 in Ref.~\onlinecite{SI}), which agrees well with the current-induced dynamics of the 2D frustrated bimeron~\cite{Zhang_PRB2020}.
The bimeron string is a dynamically stable only state, which shows certain layer-dependent deformation [Fig~\ref{FIG2}(e)].
The total energy increases to a stable value during the current application [Fig.~\ref{FIG3}(e)], indicating the bimeron string is an excited state maintained by $\tau_{\text{d}}$.
The numerically calculated topological charge of each FM layer only slightly varies during the transformation from the skyrmion string to the bimeron string [Fig.~\ref{FIG3}(f)], which implies that the transformation between a skyrmion string and a bimeron string is guaranteed by the topological conservation principle.
Note that the topological charge has been calibrated by slightly shifting the curve vertically, which ensures an integer charge of relaxed state.

When the current is switched off at $t=1000$ ps, the bimeron string stops rotating and spontaneously transforms back to a Bloch-type skyrmion string [Fig.~\ref{FIG2}(c)].
During this process, the system evolves back to an energetically favored perpendicularly magnetized configuration due to the effect of PMA [Figs.~\ref{FIG3}(d) and~\ref{FIG3}(e)], and the topological charge shows more obvious damped oscillation [Fig.~\ref{FIG3}(f)].
Such a phenomenon suggests that the topological spin textures can be very robust solutions in a stack of coupled FM layers, either with perpendicularly magnetized or in-plane magnetized background.

In addition, we study the $\alpha$-dependent linear motion of a Bloch-type skyrmion string at a relatively smaller $j$ as well as the $\alpha$-dependent rotation of a bimeron string at a relatively larger $j$.
The Bloch-type skyrmion string velocity and its corresponding skyrmion Hall angle decrease with increasing $\alpha$ [Fig.~\ref{FIG3}(g)].
The rotation frequency of the bimeron string is found to decrease with increasing $\alpha$ [Fig.~\ref{FIG3}(h)].


\textit{Conclusion.}
In conclusion, we have studied the static structures of N{\'e}el-type and Bloch-type skyrmion strings formed by $11$ aligned stacks of 2D frustrated pancake skyrmions.
The Bloch-type skyrmion strings with $\eta\sim\pi/2,3\pi/2$ are metastable states, which shows slightly varied $\eta$ in the $z$ dimension.
Their N{\'e}el-type counterparts with $\eta=0,\pi$ are unstable states due to the effect of DDI.
Both the Bloch-type and N{\'e}el-type skyrmion strings have layer-dependent interlayer exchange coupling energy.
For the dynamics, the Bloch-type skyrmion string shows translational motion at a small current, and it is transformed to a rotating bimeron string at a large current.
The bimeron string spontaneously transforms back to a skyrmion string when the current is switched off.

Our results reveal unusual static and dynamic properties of 3D topological spin textures in frustrated magnetic systems.
The transformation between merons and skyrmions in a chiral magnet induced by the magnetic field has been realized in experiments~\cite{Yu_Nature2018}. Future experimental explorations on the current-induced mutual transformation between the skyrmion string and the bimeron string are important for the construction of an electrically controlled multistate information storage device~\cite{Hou_NC2020} based on different 3D topological spin textures.
Possible future directions that one can explore also include the effect of a tilting field~\cite{Lin_PRB2015} on the 3D skyrmion and bimeron strings, a system with a lattice of 3D skyrmion or bimeron strings, and a system with decoupled layers.

\begin{acknowledgments}
\textit{Acknowledgments.}
X.Z. acknowledges support as an International Research Fellow of the Japan Society for the Promotion of Science (JSPS). X.Z. was supported by JSPS KAKENHI (Grant No. JP20F20363). J.X. acknowledges support from the National Natural Science Foundation of China (Grant No. 12104327). O.A.T. acknowledges support from the Australian Research Council (Grant No. DP200101027), Russian Science Foundation (Project No. 21-79-20186), NCMAS grant, and the Cooperative Research Project Program at the Research Institute of Electrical Communication, Tohoku University. G.Z. acknowledges support from the National Natural Science Foundation of China (Grants No. 51771127, No. 52171188, and No. 52111530143). J.Y. acknowledges support by the National Natural Science Foundation of China (Grant No. 51731001). Y.Z. acknowledges support from the Guangdong Special Support Project (Grant No. 2019BT02X030), Shenzhen Fundamental Research Fund (Grant No. JCYJ20210324120213037), Shenzhen Peacock Group Plan (Grant No. KQTD20180413181702403), Pearl River Recruitment Program of Talents (Grant No. 2017GC010293), and National Natural Science Foundation of China (Grants No. 11974298 and No. 61961136006). M.E. acknowledges support from the Grants-in-Aid for Scientific Research from JSPS KAKENHI (Grants No. JP18H03676 and No. JP17K05490) and support from CREST, JST (Grants No. JPMJCR20T2 and No. JPMJCR16F1). X.L. acknowledges support from the Grants-in-Aid for Scientific Research from JSPS KAKENHI (Grants No. JP20F20363, No. JP21H01364, and No. JP21K18872).
\end{acknowledgments}




\begin{thebibliography}{99}


\bibitem{Roszler_NATURE2006} U. K. R{\"o}{\ss}ler, A. N. Bogdanov, and C. Pfleiderer, Nature \textbf{442}, 797 (2006).

\bibitem{Muhlbauer_SCIENCE2009} S. M\"{u}hlbauer, B. Binz, F. Jonietz, C. Pfleiderer, A. Rosch, A. Neubauer, R. Georgii, and P. B\"{o}ni, Science \textbf{323}, 915 (2009).

\bibitem{Yu_Nature2010} X. Z. Yu, Y. Onose, N. Kanazawa, J. H. Park, J. H. Han, Y. Matsui, N. Nagaosa, and Y. Tokura, Nature \textbf{465}, 901 (2010).

\bibitem{Nagaosa_NNANO2013} N. Nagaosa and Y. Tokura, Nat. Nanotech. \textbf{8}, 899 (2013).

\bibitem{Mochizuki_Review} M.Mochizuki and S. Seki, J. Phys.: Condens. Matter \textbf{27}, 503001 (2015).

\bibitem{Wiesendanger_Review2016} R. Wiesendanger, Nat. Rev. Mat. \textbf{1}, 16044 (2016).

\bibitem{Finocchio_JPD2016} G. Finocchio, F. B{\"u}ttner, R. Tomasello, M. Carpentieri, and M. Kl{\"a}ui, J. Phys. D: Appl. Phys. \textbf{49}, 423001 (2016).

\bibitem{Kang_PIEEE2016} W. Kang, Y. Huang, X. Zhang, Y. Zhou, and W. Zhao, Proc. IEEE \textbf{104}, 2040 (2016).

\bibitem{Kanazawa_AM2017} N. Kanazawa, S. Seki, and Y. Tokura, Adv. Mater. \textbf{29}, 1603227 (2017).

\bibitem{Wanjun_PHYSREP2017} W. Jiang, G. Chen, K. Liu, J. Zang, S. G. Velthuiste, and A. Hoffmann, Phys. Rep. \textbf{704}, 1 (2017).

\bibitem{Fert_NATREVMAT2017} A. Fert, N. Reyren, and V. Cros, Nat. Rev. Mater. \textbf{2}, 17031 (2017).

\bibitem{Everschor_JAP2018} K. Everschor-Sitte, J. Masell, R. M. Reeve, and M. Kl\"{a}ui, J. Appl. Phys. \textbf{124}, 240901 (2018).

\bibitem{Zhou_NSR2018} Y. Zhou, Natl. Sci. Rev. \textbf{6}, 210 (2019).

\bibitem{Zhang_JPCM2020} X. Zhang, Y. Zhou, K. M. Song, T.-E. Park, J. Xia, M. Ezawa, X. Liu, W. Zhao, G. Zhao, and S. Woo, J. Phys. Condens. Matter \textbf{32}, 143001 (2020).

\bibitem{Gobel_PP2021} B. G\"{o}bel, I. Mertig, and O. A. Tretiakov Phys. Rep. \textbf{895}, 1 (2021).


\bibitem{Okubo_PRL2012} T. Okubo, S. Chung, and H. Kawamura, Phys. Rev. Lett. \textbf{108}, 017206 (2012).

\bibitem{Leonov_NCOMMS2015} A. O. Leonov and M. Mostovoy, Nat. Commun. \textbf{6}, 8275 (2015).

\bibitem{Lin_PRB2016A} S.-Z. Lin and S. Hayami, Phys. Rev. B \textbf{93}, 064430 (2016).

\bibitem{Hayami_PRB2016A} S. Hayami, S.-Z. Lin, and C. D. Batista, Phys. Rev. B \textbf{93}, 184413 (2016).

\bibitem{Rozsa_PRL2016} L. R{\'o}zsa, A. De{\'a}k, E. Simon, R. Yanes, L. Udvardi, L. Szunyogh, and U. Nowak, Phys. Rev. Lett. \textbf{117}, 157205 (2016).

\bibitem{Batista_2016} C. D. Batista, S.-Z. Lin, S. Hayami, and Y. Kamiya, Rep. Prog. Phys. \textbf{79}, 084504 (2016).

\bibitem{Leonov_NCOMMS2017} A. O. Leonov and M. Mostovoy, Nat. Commun. \textbf{8}, 14394 (2017).

\bibitem{Kharkov_PRL2017} Y. A. Kharkov, O. P. Sushkov, and M. Mostovoy, Phys. Rev. Lett. \textbf{119}, 207201 (2017).

\bibitem{Xichao_NCOMMS2017} X. Zhang, J. Xia, Y. Zhou, X. Liu, H. Zhang, and M. Ezawa, Nat. Commun. \textbf{8}, 1717 (2017).

\bibitem{Yuan_PRB2017} H. Y. Yuan, O. Gomonay, and M. Kl{\"a}ui, Phys. Rev. B \textbf{96}, 134415 (2017).

\bibitem{Hou_AM2017} Z. Hou, W. Ren, B. Ding, G. Xu, Y. Wang, B. Yang, Q. Zhang, Y. Zhang, E. Liu, F. Xu, W. Wang, G. Wu, X. Zhang, B. Shen, and Z. Zhang, Adv. Mater. \textbf{29}, 1701144 (2017).

\bibitem{Hu_SR2017} Y. Hu, X. Chi, X. Li, Y. Liu, and A. Du, Sci. Rep. \textbf{7}, 16079 (2017).

\bibitem{Malottki_SR2017} S. von Malottki, B. Dupe, P. F. Bessarab, A. Delin, and S. Heinze, Sci. Rep. \textbf{7}, 12299 (2017).

\bibitem{Liang_NJP2018} J. J. Liang, J. H. Yu, J. Chen, M. H. Qin, M. Zeng, X. B. Lu, X. S. Gao, and J. Liu, New J Phys. \textbf{20}, 053037 (2018).

\bibitem{Ritzmann_NE2018} U. Ritzmann, S. von Malottki, J.-V. Kim, S. Heinze, J. Sinova, and B. Dup{\'e}, Nat. Electron. \textbf{1}, 451 (2018).

\bibitem{Kurumaji_SCIENCE2019} T. Kurumaji, T. Nakajima, M. Hirschberger, A. Kikkawa, Y. Yamasaki, H. Sagayama, H. Nakao, Y. Taguchi, T.-H. Arima, and Y. Tokura, Science \textbf{365}, 914 (2019).

\bibitem{Desplat_PRB2019} L. Desplat, J. V. Kim, and R. L. Stamps, Phys. Rev. B \textbf{99}, 174409 (2019).

\bibitem{Xia_PRApplied2019} J. Xia, X. Zhang, M. Ezawa, Z. Hou, W. Wang, X. Liu, and Y. Zhou, Phys. Rev. Applied \textbf{11}, 044046 (2019).

\bibitem{Zarzuela_PRB2019} R. Zarzuela, H. Ochoa, and Y. Tserkovnyak, Phys. Rev. B \textbf{100}, 054426 (2019).

\bibitem{Lohani_PRX2019} V. Lohani, C. Hickey, J. Masell, and A. Rosch, Phys. Rev. X \textbf{9}, 041063 (2019).

\bibitem{Gobel_PRB2019} B. G{\"o}bel, A. Mook, J. Henk, I. Mertig, and O. A. Tretiakov, Phys. Rev. B \textbf{99}, 060407(R) (2019).

\bibitem{Diep_PRB2019} I. F. Sharafullin, M. Kh. Kharrasov, and H. T. Diep, Phys. Rev. B \textbf{99}, 214420 (2019).

\bibitem{Diep_Symmetry2020} I. F. Sharafullin and H. T. Diep, Symmetry \textbf{12}, 26 (2020).

\bibitem{Diep_Entropy2019} H. T. Diep, Entropy \textbf{21}, 175 (2019).

\bibitem{Zhang_PRB2020} X. Zhang, J. Xia, L. Shen, M. Ezawa, O. A. Tretiakov, G. Zhao, X. Liu, and Y. Zhou, Phys. Rev. B \textbf{101}, 144435 (2020).

\bibitem{Xia_APL2020} J. Xia, X. Zhang, M. Ezawa, O. A. Tretiakov, Z. Hou, W. Wang, G. Zhao, X. Liu, H. T. Diep, and Y. Zhou, Appl. Phys. Lett. \textbf{117}, 012403 (2020).

\bibitem{Zhang_APL2021} X. Zhang, J. Xia, M. Ezawa, O. A. Tretiakov, H. T. Diep, G. Zhao, X. Liu, and Y. Zhou, Appl. Phys. Lett. \textbf{118}, 052411 (2021).

\bibitem{Psaroudaki_PRL2021} C. Psaroudaki and C. Panagopoulos, Phys. Rev. Lett. \textbf{127}, 067201 (2021).

\bibitem{Sinova_2021A} R. Zarzuela, D. Hill, J. Sinova, and Y. Tserkovnyak, Phys. Rev. B \textbf{103}, 174424 (2021).

\bibitem{Sinova_2021B} R. Zarzuela and J. Sinova, arXiv:2107.13330 (2021).



\bibitem{Wanjun_NPHYS2017} W. Jiang, X. Zhang, G. Yu, W. Zhang, X. Wang, M. B. Jungfleisch, J. E. Pearson, X. Cheng, O. Heinonen, K. L. Wang, Y. Zhou, A. Hoffmann, and S. G. E. Velthuiste, Nat. Phys. \textbf{13}, 162 (2017).

\bibitem{Litzius_NPHYS2017} K. Litzius, I. Lemesh, B. Kruger, P. Bassirian, L. Caretta, K. Richter, F. Buttner, K. Sato, O. A. Tretiakov, J. Forster, R. M. Reeve, M. Weigand, I. Bykova, H. Stoll, G. Schutz, G. S. D. Beach, and M. Klaui, Nat. Phys. \textbf{13}, 170 (2017).


\bibitem{Zhang_PRB2016} X. Zhang, Y. Zhou, and M. Ezawa, Phys. Rev. B \textbf{93}, 024415 (2016).

\bibitem{Sinova_2020} R. Zarzuela, V. K. Bharadwaj, K.-W. Kim, J. Sinova, and K. Everschor-Sitte, Phys. Rev. B \textbf{101}, 054405 (2020).

\bibitem{Sinova_2017} K. Everschor-Sitte, M. Sitte, T. Valet, A. Abanov, and J. Sinova, New J. Phys. \textbf{19}, 092001 (2017).



\bibitem{Clem_1991} J. R. Clem, Phys. Rev. B \textbf{43}, 7837 (1991).

\bibitem{Clem_2004} J. R. Clem, J. Supercond. Novel Magn. \textbf{17}, 613 (2004).

\bibitem{Reichhardt_PRB2001} C. J. Olson, C. Reichhardt, and V. M. Vinokur, Phys. Rev. B \textbf{64}, 140502(R) (2001).

\bibitem{Liu_APL2020} Z.-L. Liu, P. Kang, Y. Zhu, L. Liu, and H. Guo, APL Mater. \textbf{8}, 061104 (2020).

\bibitem{Reichhardt_2021} C. Reichhardt, C. J. O. Reichhardt, and M. V. Milosevic, arXiv:2102.10464.

\bibitem{Lin_PRL2018} S.-Z. Lin and C. D. Batista, Phys. Rev. Lett. \textbf{120}, 077202 (2018).

\bibitem{Reichhardt_PRB2011} C. Reichhardt and C. J. O. Reichhardt, Phys. Rev. B \textbf{84}, 174208 (2011).

\bibitem{Xia_PRB2021} J. Xia, X. Zhang, K. Mak, M. Ezawa, O. A. Tretiakov, Y. Zhou, G. Zhao, and X. Liu, Phys. Rev. B \textbf{103}, 174408 (2021).


\bibitem{Kaul_2004} E. E. Kaul, H. Rosner, N. Shannon, R. V. Shpanchenko, and C. Geibel, J. Magn. Magn. Mater. \textbf{272-276}, 922 (2004).

\bibitem{Sinova_2015} J. Sinova, S. O. Valenzuela, J. Wunderlich, C. H. Back, and T. Jungwirth, Rev. Mod. Phys. \textbf{87}, 1213 (2015).



\bibitem{OOMMF} M. J. Donahue and D. G. Porter, ``OOMMF User's Guide, Version 1.0'', Interagency Report NO. NISTIR 6376 (National Institute of Standards and Technology, Gaithersburg, MD, 1999) [http://math.nist.gov/oommf/].


\bibitem{SI} See Supplemental Material at [URL] for more information regarding the parameter dependency diagrams and supplemental videos showing the current-induced dynamics.


\bibitem{Yu_Nature2018} X. Z. Yu, W. Koshibae, Y. Tokunaga, K. Shibata, Y. Taguchi, N. Nagaosa, and Y. Tokura, Nature \textbf{564}, 95 (2018).


\bibitem{Hou_NC2020} Y. Wang, L. Wang, J. Xia, Z. Lai, G. Tian, X. Zhang, Z. Hou, X. Gao, W. Mi, C. Feng, M. Zeng, G. Zhou, G. Yu, G. Wu, Y. Zhou, W. Wang, X. Zhang, and J. Liu, Nat. Commun. \textbf{11}, 3577 (2020).


\bibitem{Lin_PRB2015} S.-Z. Lin and A. Saxena, Phys. Rev. B \textbf{92}, 180401(R) (2015).

\end{thebibliography}
\end{document}